\documentclass[a4paper,USenglish]{lipics-v2018}

\usepackage[utf8]{inputenc}
\usepackage[T1]{fontenc}
\usepackage[ruled,linesnumbered]{algorithm2e}
\usepackage{pgfplots}
\usepackage{microtype}
\usepackage{xspace}
\usepackage{csquotes}
\usepackage{listings}
\title{Restructuring expression dags for efficient parallelization}
\author{Martin Wilhelm}{Institut für Simulation und Graphik, Otto-von-Guericke-Universität~Magdeburg\\{[Universitätsplatz 2, D-39106 Magdeburg, Germany]}}{martin.wilhelm@ovgu.de}{}{}
\authorrunning{M. Wilhelm}
\Copyright{Martin Wilhelm}

\subjclass{\ \\
\\
\\
}

\EventEditors{Gianlorenzo D'Angelo}
\EventNoEds{1}
\EventLongTitle{17th International Symposium on Experimental Algorithms (SEA 2018)}
\EventShortTitle{SEA 2018}
\EventAcronym{SEA}
\EventYear{2018}
\EventDate{June 27--29, 2017}
\EventLocation{L'Aquila, Italy}
\EventLogo{}
\SeriesVolume{103}
\ArticleNo{20}

\nolinenumbers


\newcommand{\RealAlgebraic}{\texttt{Real\_algebraic}\xspace}
\newcommand{\AMBalancing}{AM-Balancing\xspace}
\newcommand{\BrLong}{Move-To-Root Restructuring\xspace}
\newcommand{\BrDesc}{Move-To-Root (MTR) Restructuring\xspace}
\newcommand{\MTR}{MTR Restructuring\xspace}
\newcommand{\PMTR}{Parameterized \MTR}
\newcommand{\BrExp}{Expression\xspace}

\newcommand{\Vop}{V_\text{op}}
\newcommand{\cpp}{C\texttt{++}}
\newcommand{\gpp}{g\texttt{++}}
\newcommand{\dualcore}{\texttt{dual\_core}\xspace}
\newcommand{\quadcore}{\texttt{quad\_core}\xspace}

\newcommand{\bre}{mtr}
\newcommand{\brem}{mtrm}
\newcommand{\bres}{mtrs}
\newcommand{\pbr}{mtr[5]}
\newcommand{\pbrm}{mtrm[5]}

\newcommand{\pbrk}{mtr[k]}
\newcommand{\pbrkm}{mtrs[k]}
\newcommand{\pbrks}{mtrm[k]}

\newcommand{\desc}{\phi}
\DeclareMathOperator{\sep}{sep}
\DeclareMathOperator{\parent}{parent}
\DeclareMathOperator{\val}{val}
\DeclareMathOperator{\vlu}{value}

\definecolor{SEQCOL1}{RGB}{227,74,51}
\definecolor{SEQCOL2}{RGB}{253,187,132}
\definecolor{SEQCOL3}{RGB}{254,232,200}

\definecolor{QCOL1}{RGB}{241,163,64}
\definecolor{QCOL2}{RGB}{153,142,195}

\begin{document}
\maketitle

\begin{abstract}
In the field of robust geometric computation it is often necessary to make exact decisions based on inexact floating-point arithmetic. One common approach is to store the computation history in an arithmetic expression dag and to re-evaluate the expression with increasing precision until an exact decision can be made.
We show that exact-decisions number types based on expression dags can be evaluated faster in practice through parallelization on multiple cores. We compare the impact of several restructuring methods for the expression dag on its running time in a parallel environment.
\end{abstract}

\section{Introduction}

In Computational Geometry, many algorithms rely on the correctness of geometric predicates, such as orientation tests or incircle tests, and may fail or produce drastically wrong output if they do not return the correct result~\cite{schirra2000}. The Exact Geometric Computation Paradigm establishes a framework for guaranteeing exact decisions based on inexact arithmetic, as it is present in real processors~\cite{yap1997}.
In accordance with this paradigm, many exact-decisions number types have been developed, such as \texttt{leda::real}~\cite{burnikel1996}, \texttt{Core::Expr}~\cite{karamcheti1999,yu2010}, \RealAlgebraic~\cite{moerig10} and \texttt{LEA}~\cite{benouamer1993}. All four named number types are based on arithmetic expression dags, i.e., they store the computation history in a directed acyclic graph and use this data structure to adaptively (re-)evaluate the expression when a decision has to be made.

While all of these number types are able to make exact decisions, they are also very slow compared to standard floating point arithmetic. Therefore continuous effort is taken to make these data types more efficient. However, none of these number types implements a strategy to take advantage of multiple cores yet.
In this work, we show that multithreading can improve the performance of expression-dag-based number types by presenting the design of a multithreaded implementation for \RealAlgebraic, as well as experimental results comparing it to its single-threaded version. 

To enable an efficient parallelization, we implement several restructuring methods for the underlying data structure and compare them with respect to their effect on multithreading. We aim for using these techniques in a general purpose number type, for which the user need not worry about implementation details.
Therefore we look specifically at situations in which restructuring increases the running time. We propose a new approach to avoid some of these situations and thus to lower the risk of worsening the performance.

\subsection{Preliminaries}
An (arithmetic) expression dag is a rooted ordered directed acyclic graph that is either
\begin{enumerate}
	\item A single node containing a number or
	\item A node representing a unary operation $(\sqrt[\leftroot{2}\uproot{2}n]{\phantom{x}}, -)$ with one, or a binary operation $(+,-,*,/)$ with two, not necessarily disjoint, arithmetic expression dags as children.
\end{enumerate}
The number type \RealAlgebraic is based on the concept of accuracy-driven computation\footnote{Also known, less accurately, by the name ``Precision-driven computation''}~\cite{yap1997}. Applying operations leads to the creation of an expression dag instead of the calculation of an approximation. When a decision has to be made, the maximum accuracy needed for the decision to be exact is determined and each node is (re)computed with a precision that is sufficient to guarantee the desired final accuracy\footnote{Actually, this is an iterative process with increasing accuracy and checks for exactness after each iteration.}.

Let $\val(E)$ be the value represented by an expression dag $E$. When an approximation for $\val(E)$ is computed, then for each node $v$ in $E$, approximations for the children of $v$, and consequently for all of its descendants, must be computed before $v$ itself can be processed. Hence the computations we have to perform are highly dependent on each other. Whether the evaluation of an expression dag can be efficiently parallelized is therefore largely determined by its structure. Generally, a shallower, more balanced structure leads to less dependencies and can be expected to facilitate a more efficient parallel evaluation.

\subsection{Related work}
Few attempts have been made to restructure arithmetic expression dags. Richard Brent showed in 1974 how to restructure arithmetic expression trees to guarantee optimal parallel evaluation time~\cite{brent74}. In 1985, Miller and Reif improved this strategy by showing how the restructuring process itself can be done in optimal time~\cite{miller85}. In our previous work, arithmetic expression dags are restructured to improve single-threaded performance by replacing tree-like substructures, containing only additions or only multiplications, by equivalent balanced trees~\cite{wilhelm17}. We call this method \emph{\AMBalancing}.

We implement a variation of Brent's approach for tree-like substructures in arithmetic expression dags and compare it with \AMBalancing.
Furthermore, we refine the algorithm based on practical observations. We do not use the strategy by Miller and Reif, since restructuring the expression dag is very cheap compared to the evaluation of bigfloat operations and therefore only minor performance gain, if any\footnote{Considerable effort would be needed to even neutralize the overhead from creating and managing different threads.}, is to be expected.

\section{Design}

In this section we briefly describe our implementation of parallel evaluation and restructuring for the dag-based number type \RealAlgebraic. A more detailed description of the parallelization can be found in the associated technical report~\cite{wilhelm18techreport}.

\subsection{Parallelization}

The running time for the evaluation of expression dags is dominated by the execution of bigfloat operations. They usually sum up to around $95\%$ of the total running time. Therefore we focus on parallelizing bigfloat operations and allow for serial preprocessing. Accuracy-driven evaluation is usually done recursively with possible re-evaluations of single nodes. For an efficient parallelization, it is necessary to eliminate recalculations to avoid expensive lock-usage. Therefore we assign the required precision to each node in a (serial) preprocessing step and afterwards evaluate the nodes in topological order as proposed by Mörig et al.~\cite{moerig15macis}.

In our approach, we assign one task to every bigfloat operation. Tasks that can be solved independently are then sorted into a task pool, where they may be executed in parallel. Each dependent task is assigned a dependency counter, which gets reduced as soon as the tasks on which they depend are finished. When the dependency counter reaches zero, the task gets sorted into the task pool. With this strategy we reduce the shared data to a minimum, such that atomic operations can be facilitated at critical points to eliminate race conditions.

The maximum number of threads working in the task pool can be adjusted, depending on the hardware available. Our tests are run with a maximum number of four threads simultaneously working on the tasks plus the main thread, which stays locked during the computation.

\subsection{Restructuring}
\label{ssc:restructuring}

In \AMBalancing, so-called \enquote{operator trees}, i.e., connected subgraphs consisting of operator nodes with only one predecessor, are replaced by equivalent balanced trees if they represent a large sum or a large product. Since sums and products are associative, this can be done without increasing the total number of operations and therefore without a large risk of significantly increasing the running time, aside from some subtleties~\cite{wilhelm17}.

In this work we extend \AMBalancing, based on the tree restructuring by Brent~\cite{brent74}. In Section~\ref{sss:brentrestructuring} we consider operator trees consisting of all basic arithmetic operations except roots and reduce their depth by continuously splitting long paths and moving the lower half to the top. In Section~\ref{sss:modbrentrestructuring} we describe a modification to this algorithm, which avoids some steps that are particularly expensive.

\subsubsection{Notation}

Let $E$ be an expression dag. We call a subgraph $T$ an \emph{operator tree} if $T$ is a maximal connected subgraph of $E$, consisting of nodes of type $\lbrace +,-,*,/,\text{- (unary)}\rbrace$ such that no node except the root of $T$ has more than one parent. We denote the set of operator nodes having one of the allowed operator types $\Vop$. We call the children of the leaves of an operator tree $T$ its \emph{operands} and let $\desc(T)$ denote their number.

An operator tree is always a tree and all operator trees in an expression $E$ are disjoint. Therefore each node $v\in\Vop$ is part of exactly one operator tree $T(v)$. For each $v\in\Vop$ we call the operands of $T(v)$ that are part of the subtree rooted at $v$ the \emph{operands of $v$} and denote their number by $\desc(v)$.
We call a path from the root of an operator tree to a leaf \emph{critical} if each node on the path has at least as many operands as its siblings.

\subsubsection{\BrLong}
\label{sss:brentrestructuring}

We briefly describe our variation of Brent's algorithm, which we refer to as \emph{\BrDesc}. Each operator tree $T$ in an expression dag $E$ is restructured separately. In each node $v$ in $T$ we store $\desc(v)$, i.e., the number of operands of $v$. Then we search for a split node $v_s$ on a critical path of $T$, such that $\desc(v_s)\leq\frac{1}{2}\desc(T)<\desc(\parent(v_s))$.

\begin{figure}[ht]
\centering
\includegraphics{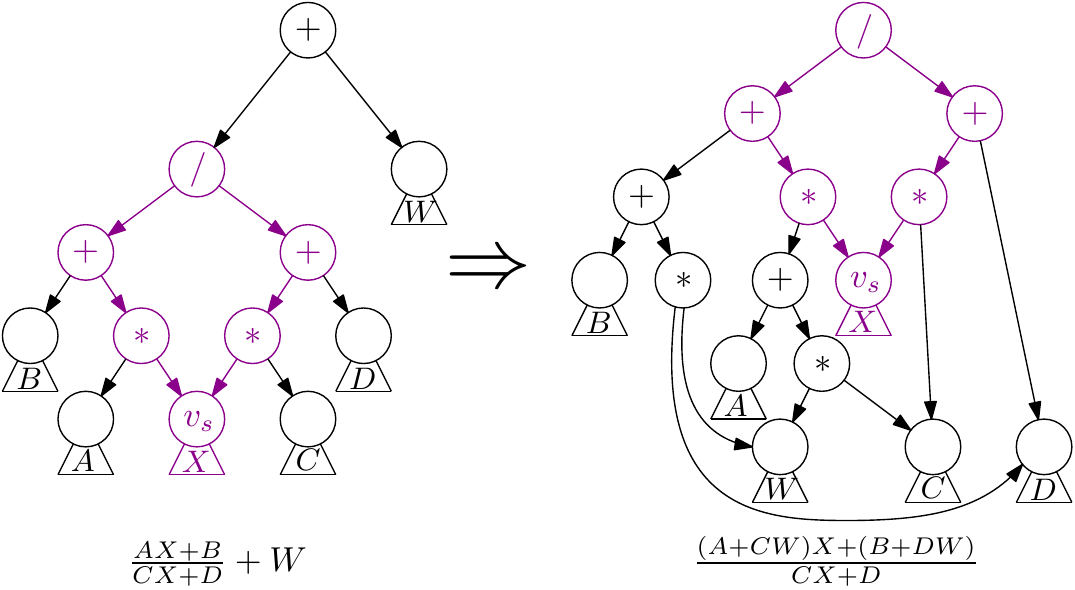}
\caption{Incorporating an addition into an expression of the form $(AX+B)/(CX+D)$. The main structure of the expression is highlighted before and after the restructuring. The expression dag grows by two additional multiplications and one additional addition due to the denominator.}
\label{fig:brentoperations}
\end{figure}

Let $X$ be the subexpression at $v_s$. We restructure $T$, such that it now represents an equivalent expression of the form
\(
\frac{AX+B}{CX+D}
\)
with (maybe trivial) subexpressions $A,B,C,D$. Starting with the expression $X$ at $v_s$, we raise $v_s$ to the top of the tree while maintaining the form 
\(
\frac{AX+B}{CX+D}
\).
We say that we \emph{incorporate} the operations along the way from $v_s$ to the root into the expression.
The restructuring needed to incorporate an addition is shown exemplarly in Figure~\ref{fig:brentoperations}.

After restructuring, the length of the path from the root to $v_s$ is reduced to a constant. The same applies to the respective root nodes for the expressions $A$, $B$, $C$ and $D$. The algorithm then recurses on these nodes, i.e., the operator trees representing $X,A,B,C,D$ are restructured.

\paragraph{Comparison to Brent's algorithm}
Like \MTR, Brent's original algorithm searches for a split node $v_s$ on a critical path and raises it to the top of the expression tree. However, it does so in a more sophisticated manner by repeatedly splitting the path from the root to $v_s$ in half and at the same time balancing the \enquote{upper half} of the tree, i.e., the part of the tree that is left when removing the subtree rooted at $v_s$. \MTR uses a simpler approach, which moves the split node to the top first and balances the remaining parts afterwards. It is easier to implement, but leads to an increased depth when subexpressions are reused, since nodes cannot be restructured as part of a larger operator tree if they have multiple parents. We chose this approach, since it is better suited to test possible improvements and can still be expected to behave similar to Brent's algorithm in many situations.

\subsubsection{Parameterization of \MTR}
\label{sss:modbrentrestructuring}

Different operations on bigfloats have different costs\footnote{In our tests on \texttt{mpfr} bigfloats, multiplications take on average $10$-$20$ times as long as subtractions or additions and divisions take $1.5$-$2$ times as long as multiplications. Interestingly, this behavior appears to be almost independent of the size of the bigfloats.}. In the algorithm introduced in the previous section, divisions in the upper half of a critical path are risen to the top. Since divisions are expensive, this is beneficial if two divisions fuse and one of them can be replaced by a multiplication. However, each addition or subtraction that is passed adds one or two expensive multiplications to the expression (cf.\ Figure~\ref{fig:brentoperations}).

\newcommand{\fname}[1]{#1}
\begin{algorithm}[!ht]
\SetKwProg{Fn}{Function}{:}{}
\SetKwFunction{Restructure}{restructure}
\SetKwFunction{Raise}{raise}
 \Fn{\Restructure{$root$}}{
 \If{$root$ can be restructured}{
 	$exp$ = \fname{raise}($root$, $root$, $0$)\\
 	set $root$ to $exp$\\
 	\fname{restructure}($exp.A$); \fname{restructure}($exp.B$); \fname{restructure}($exp.C$); \fname{restructure}($exp.D$); \fname{restructure}($exp.X$)}
 }
 \DontPrintSemicolon\;
 \Fn{\Raise{$node$, $root$, $counter$}}{
  Create new \BrExp $exp$\\ 
  \eIf{$\desc(node) \leq \desc(root)/2$ \textbf{\upshape or} $node$ is division and $counter > THRESHOLD$}{
  	$exp$.init($node$)
  }{
 	\If{$node$ is addition or subtraction}{$counter{++}$}
 	\If{$node$ is division}{$counter = 0$}
 	\eIf{$node$ has no right child \textbf{\upshape or} $\desc(node.left) \geq \desc(node.right)$}{$exp=$\fname{raise}($node.left$,$root$,$counter$)}{$exp=$\fname{raise}($node.right$,$root$,$counter$)}
 	$exp$.incorporate($node$)
  }
  return $exp$
 }
\caption{The main part of \PMTR. A counter for the number of additions and subtractions above the current node is maintained, which may lead to a split at division nodes. If the current node is a split node, an expression of the form $(AX+B)/(CX+D)$ gets initialized. Otherwise the recursion continues and the node gets incorporated into the expression of the child (cf.\ Figure~\ref{fig:brentoperations}). Finally the root node (i.e.\ the operator tree) gets replaced by the new expression dag and the subexpressions get restructured.} 
\label{alg:modbrent}
\end{algorithm}

If this affects a large number of additions and subtractions, the benefit of raising the division vanishes. If the number of cores and therefore the expected gain from an optimal parallelization is small, it might be worthwile to allow for an increased depth to save multiplications.
Our modified algorithm works the same as the algorithm described in Section~\ref{sss:brentrestructuring}, except it counts the number of additions and subtractions along a critical path and splits at a division node if their number surpasses a certain threshold, even if the division node still contains more than half of the operands of the operator tree. If a division node is passed while the counter is still smaller than the threshold, the counter is reset to zero, since then the additions and subtractions above the division node cause additional multiplications anyway. We refer to this strategy as \emph{\PMTR}. A sketch of the main method is shown in Algorithm~\ref{alg:modbrent}.

In the worst case, we may split at a linear number of division nodes if between two division nodes are just enough additions or subtractions to pass the threshold. Then the height of the operator tree, and therefore the running time with arbitrarily many processors, grows by a linear amount. However, each split increases the height of the tree only by a constant.
We expect a similar strategy to be applicable to Brent's original algorithm, where divisions are risen to the top as well. Since a more complex counting procedure would be necessary to adapt the parameterization, this hypothesis is not evaluated in more detail in this paper.

\section{Experiments}

We perform several experiments on a dual core machine (named \dualcore) with an Intel Core i5 660, 8GB RAM and a quad core machine (named \quadcore) with an Intel Core i7-4700MQ, 16GB RAM.
For \RealAlgebraic we use Boost interval arithmetic as floating-point-filter and MPFR bigfloats for the bigfloat arithmetic. The code is compiled on Ubuntu 17.10 using \gpp~7.2.0 with \cpp11 on optimization level \texttt{O3} and linked against Boost~1.62.0 and MPFR~3.1.0. 
Test results are averaged over 25 runs each. The variance for each data point is small (the total range is usually in $\pm5\%$ of its value).
In each experiment the real value of the respective expression is computed to $|q|=50000$ binary places.

We analyze four different restructuring strategies: No restructuring (\texttt{def}), \AMBalancing (\texttt{amb}), \MTR (\texttt{\bre}) and \PMTR with a threshold of five (\texttt{\pbr}). In all of the four strategies the evaluation is done in topological order to avoid distortion of the results. For each strategy we compare the results with and without multithreading (\texttt{m}).

\subsection{Binomial coefficient}

The \AMBalancing method is particularly effective if the expression contains large sums or large products. It was conjectured that applying this restructuring method makes an expression dag more suitable for a parallel evaluation. We calculate the generalized binomial coefficient
\[
\binom{\sqrt{13}}{n} = \frac{\sqrt{13}(\sqrt{13}-1)\cdots(\sqrt{13}-n+1)}{n(n-1)\cdots 1}
\]
iteratively as in the \AMBalancing paper (cf.\ \cite{wilhelm17}).

\begin{lstlisting}
template <class NT> void bin_coeff(const int n, const long q){
	NT b = sqrt(NT(13)); NT num = NT(1); NT denom = NT(1);
	for (int i = 0; i < n; ++i) { num *= b - NT(i); denom *= NT(i+1); }
	NT bc = num/denom;
	bc.guarantee_absolute_error_two_to(q);
}
\end{lstlisting}

In this method, both the numerator and the denominator of \texttt{bc} are large, sequentially computed products. Both of them can be balanced without adding additional operations because of the associativity of the multiplication. We run \texttt{bin\_coeff} in our test environment.

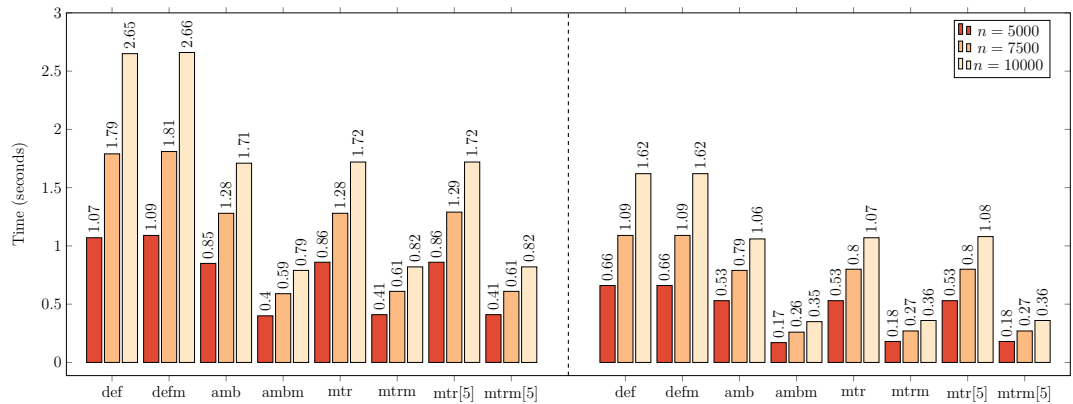
\begin{figure}[ht]
\begin{tikzpicture}[scale=0.5]
\begin{axis}[
	symbolic x coords={def,defm,amb,ambm,bre,brem,br2,br2m,placeholder,qdef,qdefm,qamb,qambm,qbre,qbrem,qbr2,qbr2m},
	xticklabels={def,defm,amb,ambm,\bre,\brem,\pbr,\pbrm,def,defm,amb,ambm,\bre,\brem,\pbr,\pbrm},
	xtick=data,
	enlarge x limits=0.05,
	ylabel=Time (seconds),
	ymax = 3,
    width  = 2\textwidth,
    height = 0.8\textwidth,
    bar width = 0.4cm,
    ybar,
    label style={font=\Large},
    tick label style={font=\Large},
    legend style={font=\Large},
    nodes near coords,
    every node near coord/.append style={rotate=90, anchor=west, 
    									 /pgf/number format/precision=4,font=\Large}
]

\addplot[fill=SEQCOL1] coordinates {(def,1.07) (defm,1.09) (amb,0.85) (ambm,0.4) (bre,0.86) (brem,0.41) (br2,0.86) (br2m,0.41) 
									(qdef,0.66) (qdefm,0.66) (qamb,0.53) (qambm,0.17) (qbre,0.53) (qbrem,0.18) (qbr2,0.53) (qbr2m,0.18)};
\addplot[fill=SEQCOL2] coordinates {(def,1.79) (defm,1.81) (amb,1.28) (ambm,0.59) (bre,1.28) (brem,0.61) (br2,1.29) (br2m,0.61)
									(qdef,1.09) (qdefm,1.09) (qamb,0.79) (qambm,0.26) (qbre,0.8) (qbrem,0.27) (qbr2,0.8) (qbr2m,0.27)};
\addplot[fill=SEQCOL3] coordinates {(def,2.65) (defm,2.66) (amb,1.71) (ambm,0.79) (bre,1.72) (brem,0.82) (br2,1.72) (br2m,0.82)
									(qdef,1.62) (qdefm,1.62) (qamb,1.06) (qambm,0.35) (qbre,1.07) (qbrem,0.36) (qbr2,1.08) (qbr2m,0.36)};

\path
  (axis cs:br2m, \pgfkeysvalueof{/pgfplots/ymin})
    -- coordinate (min)
  (axis cs:qdef, \pgfkeysvalueof{/pgfplots/ymin})
  (axis cs:br2m, \pgfkeysvalueof{/pgfplots/ymax})
    -- coordinate (max)
  (axis cs:qdef, \pgfkeysvalueof{/pgfplots/ymax});

\draw[thick, dashed] (min) -- (max);

\legend{$n=5000$,$n=7500$,$n=10000$}
\end{axis}
\end{tikzpicture}

\caption{The results of the binomial coefficient test for \dualcore (left) and \quadcore (right). The original structure of the expression dag is not suited for a parallel evaluation. \AMBalancing leads to a beneficial structure. \MTR and \PMTR have a similar effect as \AMBalancing.}
\label{fig:bincoeff}
\end{figure}

The results are shown in Figure~\ref{fig:bincoeff}. Switching to multithreading while retaining the structure of the expression dag does not have a positive effect on the performance, since the operator nodes are highly dependent on each other. Applying \AMBalancing does not only directly increase the performance, but also makes the structure much more favorable for parallel evaluation. On the dual core machine the maximal possible performance gain is achieved. With a quad core we still get an improvement, but only by a factor of about $2.8$.

For large additions and large multiplications, \MTR behaves similar to \AMBalancing in the sense that it builds an (almost) balanced tree. So, unsurprisingly, the running times for \MTR and \PMTR closely resemble the results for \AMBalancing.

\subsection{Random operations}
\label{ssc:randomoperations}

A different behavior of the restructuring methods is to be expected if algorithms use many different operators in varying order. We simulate this behavior by performing random operations on an expression.

\begin{lstlisting}
template <class NT> void random_operations(const int n, const long q, const int FADD = 1, const int FSUB = 1, const int FMUL = 1, const int FDIV = 1){
	std::random_device rd; std::mt19937 gen(rd());
	std::uniform_int_distribution<> idis(0,FADD+FSUB+FMUL+FDIV-1);
	std::exponential_distribution<> rdis(1);

	double r; NT result = NT(1);
	for (int i = 0; i < n; ++i) {
		const int nbr = idis(gen);
		do { r = rdis(gen); } while (r == 0);

		if (nbr < FADD) result += sqrt(NT(r));
		else if (nbr < FADD+FSUB) result -= sqrt(NT(r));
		else if (nbr < FADD+FSUB+FMUL) result *= sqrt(NT(r));
		else result /= sqrt(NT(r));
	}
	result.guarantee_absolute_error_two_to(q);
}
\end{lstlisting}

We exploit two kinds of randomness in this test. First, we randomly choose one of the operators $\lbrace +,-,*,/\rbrace$. The parameters \texttt{FADD}, \texttt{FSUB}, \texttt{FMUL} and \texttt{FDIV} determine their respective fractions of the total number of operators. If all of them are set to one, the operations are equally distributed. Second, we randomly choose a real positive number on which to apply the operator. We use an exponential distribution with a mean of one instead of a uniform distribution. \RealAlgebraic behaves differently for very large or very small numbers. In a uniform distribution most of the random numbers are larger than one. Therefore through repeated multiplication (division) numbers get very large (very small). Since we want to be able to compare the actual costs of multiplications and divisions, we have to avoid distorting the results through side effects of the experiment.

In our test we use the square roots of the random floating point numbers we get to generate numbers with an infinite floating point representation. When performing a division, \RealAlgebraic must check whether the denominator is zero. So it must decide at which point the established accuracy is sufficient to guarantee that the value of an expression is zero. This decision is made by comparing the current error bound to a separation bound, which is a number $\sep(X)$ for an expression $X$, for which $|\vlu(X)|>0\Rightarrow|\vlu(X)|>\sep(X)$.
The separation bound we use is a variation of the separation bound by Burnikel et al.~\cite{burnikel2009,schmitt2004}. It relies on having a meaningful bound for the algebraic degree of an expression, which we cannot provide in an expression with a lot of square roots. However, since we know that none of the denominators we get during our experiments can become zero, we can safely set the separation bound to zero and stop computing as soon as we can separate our denominator from zero with an error bound.

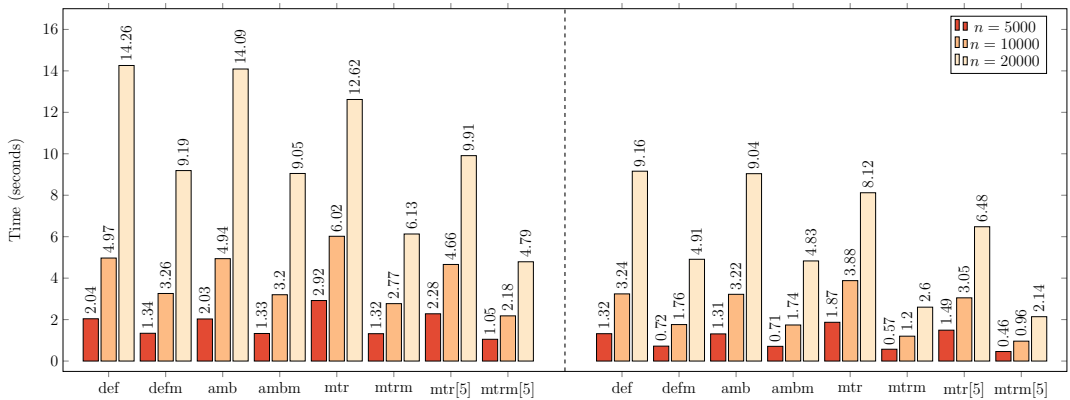
\begin{figure}[ht]
\begin{tikzpicture}[scale=0.5]
\begin{axis}[
	symbolic x coords={def,defm,amb,ambm,bre,brem,br2,br2m,placeholder,qdef,qdefm,qamb,qambm,qbre,qbrem,qbr2,qbr2m},
	xticklabels={def,defm,amb,ambm,\bre,\brem,\pbr,\pbrm,def,defm,amb,ambm,\bre,\brem,\pbr,\pbrm},
	xtick=data,
	enlarge x limits=0.05,
	ylabel=Time (seconds),
	ymin = -0.5,
	ymax = 17,
    width  = 2\textwidth,
    height = 0.8\textwidth,
    bar width = 0.4cm,
    ybar,
    label style={font=\Large},
    tick label style={font=\Large},
    legend style={font=\Large},
    nodes near coords,
    every node near coord/.append style={rotate=90, anchor=west, 
    									 /pgf/number format/precision=4,font=\Large}
]
\addplot[fill=SEQCOL1] coordinates {(def,2.04) (defm,1.34) (amb,2.03) (ambm,1.33) (bre,2.92) (brem,1.32) (br2,2.28) (br2m,1.05)
									(qdef,1.32) (qdefm,0.72) (qamb,1.31) (qambm,0.71) (qbre,1.87) (qbrem,0.57) (qbr2,1.49) (qbr2m,0.46) };
\addplot[fill=SEQCOL2] coordinates {(def,4.97) (defm,3.26) (amb,4.94) (ambm,3.2) (bre,6.02) (brem,2.77) (br2,4.66) (br2m,2.18)
									(qdef,3.24) (qdefm,1.76) (qamb,3.22) (qambm,1.74) (qbre,3.88) (qbrem,1.2) (qbr2,3.05) (qbr2m,0.96) };
\addplot[fill=SEQCOL3] coordinates {(def,14.26) (defm,9.19) (amb,14.09) (ambm,9.05) (bre,12.62) (brem,6.13) (br2,9.91) (br2m,4.79)
									(qdef,9.16) (qdefm,4.91) (qamb,9.04) (qambm,4.83) (qbre,8.12) (qbrem,2.6) (qbr2,6.48) (qbr2m,2.14) };
	
\path
  (axis cs:br2m, \pgfkeysvalueof{/pgfplots/ymin})
    -- coordinate (min)
  (axis cs:qdef, \pgfkeysvalueof{/pgfplots/ymin})
  (axis cs:br2m, \pgfkeysvalueof{/pgfplots/ymax})
    -- coordinate (max)
  (axis cs:qdef, \pgfkeysvalueof{/pgfplots/ymax});

\draw[thick, dashed] (min) -- (max);	

\legend{$n=5000$,$n=10000$,$n=20000$}
\end{axis}
\end{tikzpicture}

\caption{Results for performing random operations uniformly distributed on \dualcore (left) and \quadcore (right). \AMBalancing has no significant effect. \MTR and \PMTR worsen the single-threaded performance for small numbers of operands, but improve the multithreaded performance. \PMTR performs better than \MTR in all cases.}
\label{fig:rduniform}
\end{figure}

The results for a uniform distribution of operators are shown in Figure~\ref{fig:rduniform}. Although the structure of the dag is largely unsuited for parallelization, we can see a significant difference between the single-threaded and the multi-threaded variant even without restructuring due to the parallel evaluation of the square root operations. \AMBalancing shows no effect at all on the overall performance.

For smaller inputs, \MTR has a negative effect on the single-threaded performance. The predominant cause of this is the propagation of divisions as described in Section~\ref{ssc:restructuring}, which results on average in about nine multiplications per processed division. \PMTR reduces this ratio to about~$5.5$ multiplications per processed division by leaving about one tenth of them unprocessed. For large inputs, \MTR has a positive effect even without multithreading due to the decrease in depth of the main operator tree and, consequently, a decrease in the maximum accuracy needed for the square root operations in its leaves (cf.~\cite{wilhelm17}).

In the case of multithreading, both approaches based on Brent's algorithm show the desired effect, increasing the speedup for the dual core from around~$1.7$ to an optimal~$2$ and for the quad core from around~$1.8$ to about~$2.8$. As a consequence, they are able to beat direct parallelization in every test case. \PMTR performs slightly better than \MTR due to the increase in single-threaded performance while maintaining the same speedup.

\subsection{Random operations with mostly divisions}
\label{ssc:mostlydivisions}

Raising divisions to the top is bad only if many additions and subtractions are passed during the procedure. If instead divisions can be combined and therefore replaced by multiplications, the overall effect is positive. We test this by shifting the operator distribution such that nine out of ten operators are divisions, i.e., by setting \texttt{FADD~=~1, FSUB~=~1, FMUL~=~1, FDIV~=~27}. Since \AMBalancing shows no differences compared to the default number type, we exclude it from further experiments.

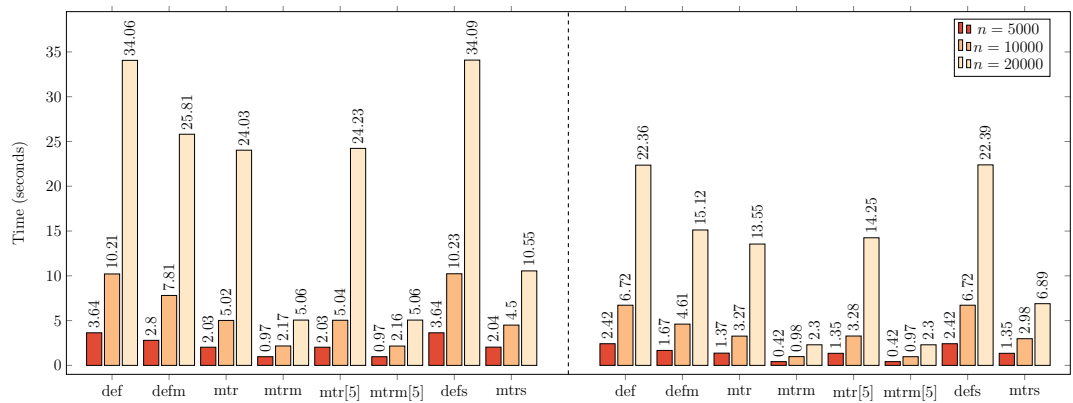
\begin{figure}[ht]
\begin{tikzpicture}[scale=0.5]
\begin{axis}[
	symbolic x coords={def,defm,bre,brem,br2,br2m,defs,bres,placeholder,qdef,qdefm,qbre,qbrem,qbr2,qbr2m,qdefs,qbres},
	xticklabels = {def,defm,\bre,\brem,\pbr,\pbrm,defs,\bres,def,defm,\bre,\brem,\pbr,\pbrm,defs,\bres},
	xtick=data,
	enlarge x limits=0.05,
	ylabel=Time (seconds),
	ymin = -1,
	ymax = 39.5,
    width  = 2\textwidth,
    height = 0.8\textwidth,
    bar width = 0.4cm,
    ybar,
    label style={font=\Large},
    tick label style={font=\Large},
    legend style={font=\Large},
    nodes near coords,
    every node near coord/.append style={rotate=90, anchor=west, 
    									 /pgf/number format/precision=4,font=\Large}
]

\addplot[fill=SEQCOL1] coordinates {(def,3.64) (defm,2.8) (bre,2.03) (brem,0.97) (br2,2.03) (br2m,0.97) (defs,3.64) (bres,2.04)
									(qdef,2.42) (qdefm,1.67) (qbre,1.37) (qbrem,0.42) (qbr2,1.35) (qbr2m,0.42) (qdefs,2.42) (qbres,1.35)};
\addplot[fill=SEQCOL2] coordinates {(def,10.21) (defm,7.81) (bre,5.02) (brem,2.17) (br2,5.04) (br2m,2.16) (defs,10.23) (bres,4.5)
									(qdef,6.72) (qdefm,4.61) (qbre,3.27) (qbrem,0.98) (qbr2,3.28) (qbr2m,0.97) (qdefs,6.72) (qbres,2.98)};
\addplot[fill=SEQCOL3] coordinates {(def,34.06) (defm,25.81) (bre,24.03) (brem,5.06) (br2,24.23) (br2m,5.06) (defs,34.09) (bres,10.55)
									(qdef,22.36) (qdefm,15.12) (qbre,13.55) (qbrem,2.3) (qbr2,14.25) (qbr2m,2.3) (qdefs,22.39) (qbres,6.89)};

\path
  (axis cs:bres, \pgfkeysvalueof{/pgfplots/ymin})
    -- coordinate (min)
  (axis cs:qdef, \pgfkeysvalueof{/pgfplots/ymin})
  (axis cs:bres, \pgfkeysvalueof{/pgfplots/ymax})
    -- coordinate (max)
  (axis cs:qdef, \pgfkeysvalueof{/pgfplots/ymax});

\draw[thick, dashed] (min) -- (max);

\legend{$n=5000$,$n=10000$,$n=20000$}
\end{axis}
\end{tikzpicture}

\caption{Test results for random operations with mostly divisions on \dualcore (left) and \quadcore (right). \MTR causes an enormous increase in performance. The large jump between single- and multithreading can be partially ascribed to a slight change in implementation that shows a positive effect on the restructured expression dag. Applying this change without multithreading leads to the results shown in \texttt{defs} and \texttt{\bres}. \PMTR behaves similar to \MTR.}
\label{fig:rdmostlydivisions}
\end{figure}

Considering the results shown in Figure~\ref{fig:rdmostlydivisions} it becomes evident that restructuring alone reduces the running time significantly in this situation. When switching to multithreading we get an even bigger improvement, which, however, can not be explained by parallelizing alone. Instead about half of the improvement stems from an implementation detail when computing the separation bound for separating denominators from zero in the multithreaded version. The change usually leads to a slight overhead, but also happens to prevent a slow-down if many checks have to be made. It is explained in detail in the associated technical report~\cite{wilhelm18techreport}.
After restructuring we have many big denominators, for which a separation bound must be computed and therefore we get an improvement. The default number type, on the other hand, does not benefit from the new separation bound computation strategy. The data points for \texttt{defs} and \texttt{\bres} represent the test results for single-threaded evaluation with the change in separation bound computation applied.

In contrast to our previous test, \PMTR does not perform better than \MTR, since there are few to none situations in which the condition for the improvement gets triggered. However, more importantly, the modified algorithm also does not perform worse than the original one.

\subsection{Random operations with few divisions}

With few divisions, compared to the number of additions and subtractions, we should expect the opposite effect from the previous experiment. \MTR should lead to an decrease in singlethreaded performance and \PMTR should perform better than \MTR. We set our input parameters to \texttt{FADD~=~3, FSUB~=~3, FMUL~=~3, FDIV~=~1}, such that only one out of ten operations is a division.

\begin{figure}[ht]
\begin{tikzpicture}[scale=0.5]
\begin{axis}[
	symbolic x coords={def,defm,bre,brem,br2,br2m,placeholder,qdef,qdefm,qbre,qbrem,qbr2,qbr2m},
	xticklabels={def,defm,\bre,\brem,\pbr,\pbrm,def,defm,\bre,\brem,\pbr,\pbrm},
	xtick=data,
	enlarge x limits=0.1,
	ylabel=Time (seconds),
	ymin = -0.5,
	ymax = 16,
    width  = 2\textwidth,
    height = 0.8\textwidth,
    bar width = 0.5cm,
    ybar,
    label style={font=\Large},
    tick label style={font=\Large},
    legend style={font=\Large},
    nodes near coords,
    every node near coord/.append style={rotate=90, anchor=west, 
    									 /pgf/number format/precision=4,font=\Large}
]
\addplot[fill=SEQCOL1] coordinates {(def,1.7) (defm,1.05) (bre,3.09) (brem,1.38) (br2,2.04) (br2m,0.99)
									(qdef,1.12) (qdefm,0.55) (qbre,1.99) (qbrem,0.59) (qbr2,1.34) (qbr2m,0.46)};
\addplot[fill=SEQCOL2] coordinates {(def,4.11) (defm,2.5) (bre,6.45) (brem,2.87) (br2,4.24) (br2m,2.11)
									(qdef,2.67) (qdefm,1.3) (qbre,4.12) (qbrem,1.23) (qbr2,2.8) (qbr2m,0.99)};
\addplot[fill=SEQCOL3] coordinates {(def,11.27) (defm,6.76) (bre,13.32) (brem,6.08) (br2,9.31) (br2m,4.76)
									(qdef,7.25) (qdefm,3.49) (qbre,8.47) (qbrem,2.83) (qbr2,6.12) (qbr2m,2.3)};

\path
  (axis cs:br2m, \pgfkeysvalueof{/pgfplots/ymin})
    -- coordinate (min)
  (axis cs:qdef, \pgfkeysvalueof{/pgfplots/ymin})
  (axis cs:br2m, \pgfkeysvalueof{/pgfplots/ymax})
    -- coordinate (max)
  (axis cs:qdef, \pgfkeysvalueof{/pgfplots/ymax});

\draw[thick, dashed] (min) -- (max);

\legend{$n=5000$,$n=10000$,$n=20000$}
\end{axis}
\end{tikzpicture}

\caption{Test results for random operations with few divisions on \dualcore (left) and \quadcore (right). \MTR worsens the single-threaded performance and is not able to outperform direct multithreading for smaller inputs. \PMTR peforms better in all tests, although having a worse speedup factor on the quad core.}
\label{fig:rdfewdivisions}
\end{figure}
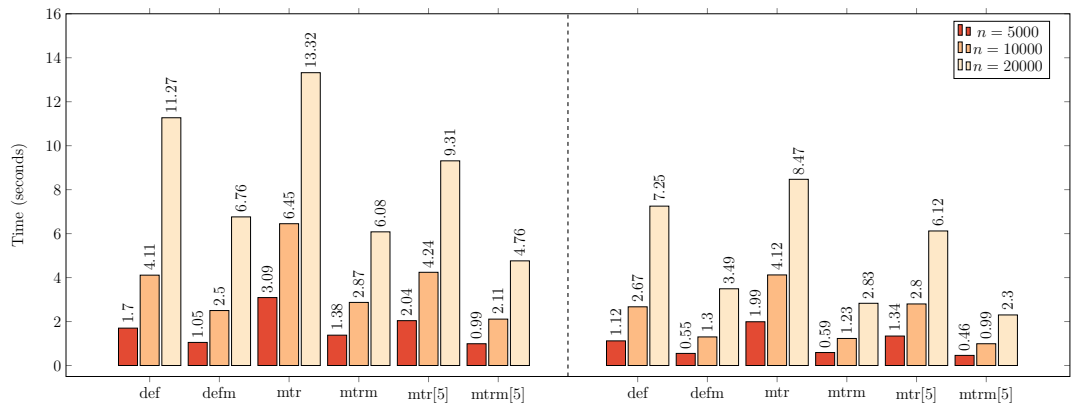

Our test results confirm these expectations (cf.\ Figure~\ref{fig:rdfewdivisions}). For small inputs, \MTR performs worse than no restructuring even in a parallel environment. \PMTR on the other hand performs better than the default in all parallel tests except for the test with the smallest number of operands on a dual core. This effect strengthens when the number of divisions further decreases.

However, it should be noted that while on a dual core the speedup through parallelization is optimal for both variants, on a quad core \MTR allows for a speedup of about~$3$, whereas \PMTR only reaches a speedup of around~$2.7$. The modified algorithm leaves about forty percent of the divisions untouched for this operator distribution, which manifests in a significant effect on the expression dag's degree of independence.

\subsection{Parameter-dependence of \PMTR}

In the experiments in the previous sections we set the threshold $k$ for \PMTR to five without an explanation. In this section we make evident that, while there is always an optimal choice for this parameter, most choices are not actually bad compared to standard \MTR.

The threshold indicates the number of additions and subtractions that are affected by incorporating a division node into the new structure and therefore sets the benefit from raising a division to the top in relation to the number of additional multiplication nodes it causes (cf.\ Section~\ref{sss:modbrentrestructuring}).
If $k=0$, restructuring never incorporates divisions in its expressions, therefore all expressions are of the form $AX+B$. For $k=1$ divisions are only incorporated if they are followed exclusively by divisions, multiplications or negations. With increasing $k$ it becomes less frequent that a division cannot be passed during restructuring, leading to \PMTR behaving more and more similar to \MTR.

\begin{figure}[hbt]
\captionsetup[subfigure]{justification=centering}
\centering
\begin{subfigure}{0.5\linewidth}
\centering
\begin{minipage}{2\linewidth}
\begin{tikzpicture}[scale=0.5]
\begin{axis}[
	xlabel=Threshold (k),
	xtick=data,
	ylabel=Time (seconds),
	ymax = 10,
    label style={font=\Large},
    tick label style={font=\Large},
    legend style={font=\Large,at={(0.5,0.97)},anchor=north},
    legend columns=-1,
    width  = 0.95\textwidth
]
\addplot[QCOL1,mark=*] coordinates { (0,8.31) (1,8.31) (2,8.31) (3,8.31) (4,8.31) (5,8.31) (6,8.31) (7,8.31) (8,8.31) (9,8.31) (10,8.31) (11,8.31) (12,8.31) (13,8.31) (14,8.31) (15,8.31) };

\addplot[QCOL2,mark=*] coordinates { (0,7.97) (1,7.08) (2,6.69) (3,6.51) (4,6.5) (5,6.57) (6,6.7) (7,6.79) (8,6.93) (9,7.09) (10,7.27) (11,7.41) (12,7.61) (13,7.74) (14,7.87) (15,8.02) };

\addplot[QCOL1,mark=square*] coordinates { (0,2.62) (1,2.62) (2,2.62) (3,2.62) (4,2.62) (5,2.62) (6,2.62) (7,2.62) (8,2.62) (9,2.62) (10,2.62) (11,2.62) (12,2.62) (13,2.62) (14,2.62) (15,2.62) };

\addplot[QCOL2,mark=square*] coordinates { (0,4.06) (1,3.18) (2,2.7) (3,2.39) (4,2.23) (5,2.14) (6,2.08) (7,2.06) (8,2.06) (9,2.08) (10,2.13) (11,2.15) (12,2.21) (13,2.26) (14,2.33) (15,2.37) };

\legend{\bre,\pbrk,\brem,\pbrkm}
\end{axis}
\end{tikzpicture}
\end{minipage}
\subcaption{Uniformly distributed operators}
\end{subfigure}%
\begin{subfigure}{0.5\linewidth}
\centering
\begin{minipage}{2\linewidth}
\begin{tikzpicture}[scale=0.5]
\begin{axis}[
	xlabel=Threshold (k),
	xtick=data,
	ylabel=Time (seconds),
	ymax = 8,
    label style={font=\Large},
    tick label style={font=\Large},
    legend style={font=\Large,at={(0.5,0.97)},anchor=north},
    legend columns=-1,
    width  = 0.95\textwidth
]
\addplot[QCOL1,mark=*] coordinates { (0,6.86) (1,6.86) (2,6.86) (3,6.86) (4,6.86) (5,6.86) (6,6.86) (7,6.86) (8,6.86) (9,6.86) (10,6.86) (11,6.86) (12,6.86) (13,6.86) (14,6.86) (15,6.86) };

\addplot[QCOL2,mark=*] coordinates { (0,6.93) (1,6.02) (2,6.46) (3,6.86) (4,6.85) (5,6.86) (6,6.86) (7,6.86) (8,6.84) (9,6.85) (10,6.87) (11,6.85) (12,6.87) (13,6.84) (14,6.87) (15,6.87) };

\addplot[QCOL1,mark=square*] coordinates { (0,2.31) (1,2.31) (2,2.31) (3,2.31) (4,2.31) (5,2.31) (6,2.31) (7,2.31) (8,2.31) (9,2.31) (10,2.31) (11,2.31) (12,2.31) (13,2.31) (14,2.31) (15,2.31) };

\addplot[QCOL2,mark=square*] coordinates { (0,2.51) (1,1.73) (2,2.01) (3,2.28) (4,2.31) (5,2.3) (6,2.29) (7,2.29) (8,2.3) (9,2.29) (10,2.31) (11,2.3) (12,2.3) (13,2.31) (14,2.32) (15,2.3) };

\legend{\bres,\pbrks,\brem,\pbrkm}
\end{axis}
\end{tikzpicture}
\end{minipage}
\subcaption{Mostly divisions (nine out of ten)}
\end{subfigure}
\caption{Comparison of different parameter choices for \PMTR with $n=20000$ random operations on \quadcore. For large values of $k$, the parameterized approach becomes increasingly similar to \MTR. For small values of $k$ the original approach is slightly better in the multithreaded version. The optimal choice for $k$ depends on the division ratio.
}
\label{fig:threshold}
\end{figure}
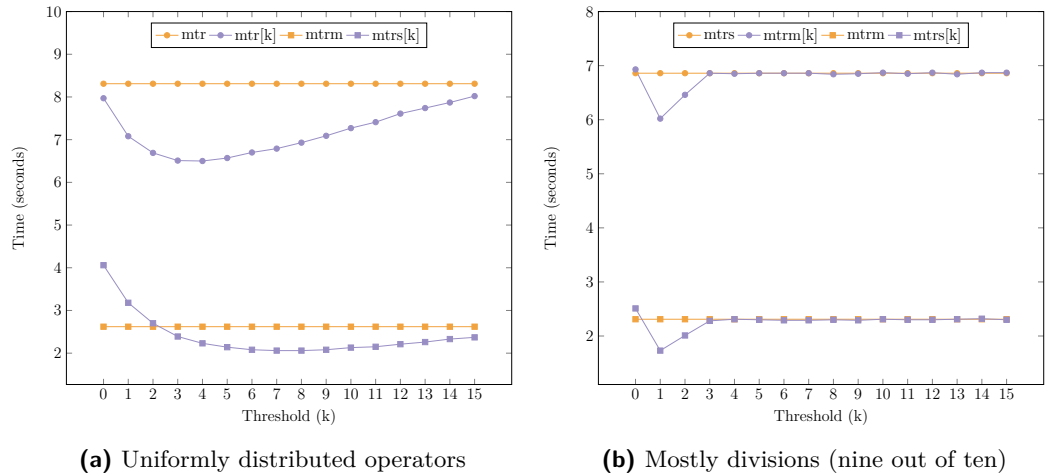

In Figure~\ref{fig:threshold} the running time for the experiments from Section~\ref{ssc:randomoperations} and Section~\ref{ssc:mostlydivisions} for different values of $k$ are shown. For the second experiment, we use the modified separation bound computation strategy to ensure comparability (cf.\ Section~\ref{ssc:mostlydivisions}). The results demonstrate that most of the choices for $k$ improve the performance of the algorithm. Also, they confirm that for high $k$ the parameterized approach is almost identical to the original algorithm.

For $k=0$ the parameterized approach performs worse in both experiments in the parallel version, although slightly increasing the single-threaded performance for uniformly distributed operators. Surprisingly, \PMTR is faster than \MTR for $k=1$ when nine out of ten operators are divisions, despite in this case restructuring tends to replace divisions by multiplications, which then can be balanced due to their associativity. However, since only one out of fifteen operations is an addition or subtraction, the loss of independence is small compared to the gain from avoiding additional multiplications. 

The optimal choice for $k$ depends on the ratio between divisions and additions/subtractions. If this ratio gets smaller, the optimal $k$ increases. At the same time, the difference between different choices for $k$ decreases. Therefore for a small ratio and smaller values of $k$ the parameterized approach still performs better than the original version.

\section{Summary}

Multithreading can be an effective tool to speed up the performance of expression-dag-based number types. Applying \MTR to expression dags allows us to benefit from multithreading even when faced with a structure with a high number of dependencies, although bearing the risk of lowering the performance. \AMBalancing can create favorable structures with low risk of worsening the end result, but is not widely applicable.
The parameterized version of \MTR lowers the risk of significant performance loss, while at the same time maintaining most of the benefits of the original algorithm. In a general purpose number type, we therefore suggest using \PMTR, with a sensible choice of $k$, if the evaluation should be done in parallel. For $k$, a small number greater than zero is advisable.

\section{Future Work}

This work addresses parallelization on multiple CPUs. While it seems unlikely that complex expression dags can be efficiently parallelized on a GPU, it may be possible to do so for the underlying bigfloats. Since bigfloat operations still constitute the most expensive part of exact-decisions number types, this may lead to a significant speedup. Furthermore in this work we only restructure tree-like subgraphs to avoid (possibly exponential) blow-up of our structure. However, with a larger number of cores it might be worthwile to split up some nodes with multiple parents to eliminate or shorten critical paths or at least weigh such nodes accordingly in the higher-level operator trees.
Finally, the performance gain due to the parameterization cannot be fully attributed to interactions between divisions and additions. It may prove useful to extend the new strategy to consider multiplications over large sums.

\bibliography{lit}
\bibliographystyle{plainurl}
\end{document}